\documentclass[prl,aps, amsfonts, nofootinbib, notitlepage]{revtex4-1}

\usepackage{caption}
\usepackage{subfig}

\usepackage{epsfig}
\usepackage{amssymb}
\usepackage{amsthm}
\usepackage{amsmath}
\usepackage{amsmath}

\newcommand{\beq}{\begin{equation}}
\newcommand{\eeq}{\end{equation}}
\newcommand{\bea}{\begin{eqnarray}}
\newcommand{\eea}{\end{eqnarray}}

\newcommand{\nn}{\nonumber}
\newcommand{\benn}{\begin{displaymath}}
\newcommand{\eenn}{\end{displaymath}}
\newcommand{\no}{\frac{2\pi n_0}{\beta}}
\newcommand{\nv}{\frac{2\pi \vec{n}}{L}}

\begin{document}

\title{Heavy Meson Masses in the $\epsilon$-Regime of HM$\chi$PT }
\author{Ra\'ul A. Brice\~no\footnote{{\tt briceno@uw.edu}}}
\affiliation{Department of Physics, University of Washington\\
Box 351560, Seattle, WA 98195, USA}

\begin{abstract}

The pseudoscalar and vector heavy meson masses are calculated in the  $\epsilon$-regime of Heavy Meson Chiral Perturbation Theory to order $\epsilon^4$. The results presented will allow the determination of low-energy coefficients (LECs) directly from Lattice QCD calculations of the heavy mesons masses for lattices that satisfy the $\epsilon$-regime criteria. In particular, the LECs that parametrize the NLO volume dependance of the heavy meson masses are necessary for evaluating the light pseudoscalar meson ($\pi$, $\text{K}$, $\eta$) and heavy meson ($\{D^0, D^+, D^+_s\}$, $\{{B}^-,\bar{B}^0,\bar{B}^0_s\}$) scattering phase shifts. 

 \end{abstract}
\maketitle

 %%%%%%%%%%%%%%%%%Lagrangian and Field Definitions
\section{Introduction}

Understanding the properties of systems composed of heavy mesons, containing a single heavy quark, and the pseudo-Goldstone bosons (pGB) of Quantum Chromodynamics (QCD) is currently a topic of high interest. This interest has been partly triggered by the renaissance of charmonium and open-charm studies. A resonance that has initiated much discussion is the narrow $\text{D}_{\text{s}0}^*(2317)$, first observed by the BaBar collaboration \cite{babar0}. This resonance couples to the S-wave DK continuum scattering state \cite{Beveren1, Beveren2}. 
At low energies the strength of the DK interaction is predominantly parametrized by the scattering length. This has resulted in several theoretical studies that have attempted to determine the S-wave scattering lengths in the pGB-heavy meson scattering channels \cite{scatteringlengths1, scatteringlengths2, scatteringlengths3, scatteringlengths4}. The determination of these scattering lengths would not only help discern the heavy meson spectrum, but is needed in order to evaluate transport coefficients of systems containing heavy-light mesonic species, e.g. the hadronic phase of heavy ion collisions.  

Currently, a combination of effective field theories (EFTs) and Lattice Quantum Chromodynamics (LQCD) provides the best option for performing reliable calculations of low energy QCD observables (reviews on these topics include \cite{theory1, theory2, lattice1, lattice2, lattice3,latlat, lattice5}). Heavy Meson Chiral Perturbation Theory (HM$\chi$PT) \cite{heavychiPT1, heavychiPT2, heavychiPT3}  is the low-energy EFT for studying strong-interaction quantities of mesons containing a single heavy quark and a single light antiquark. The non-perturbative QCD contributions to HM$\chi$PT are parametrized by low-energy coefficients (LECs).  
The predictive power of HM$\chi$PT is currently limited by the poor determination of these LECs, e.g. currently next-to-leading order (NLO) LECs are determined within a factor of three of precision, resulting in scattering lengths that are known within a factor of three \cite{scatteringlengths3, scatteringlengths4}. The results outlined in the work will help reduce the uncertainties of LECs needed in the evaluation of pGB-heavy meson scattering. 

Historically, LQCD calculations have used moderate volumes and unphysically large pion masses. With advances in computing technology, 
performing LQCD calculation at the physical point ($m_{\pi}\approx 140$ {MeV}) of QCD is now a reality. 
However, limited computer resources require state of the art calculation with physical pion masses to be performed with small physical volumes. This leads to sizable volume effects  contributing to the quantities of interest, and while it is natural to want to remove them, 
these effects can hold physically important information. More specifically, volume effects are parametrized by the LECs of the EFT, therefore by evaluating physical observables in small volumes one can determine the LECs. 

In an infinite volume, the expansion parameters of HM$\chi$PT are $p/\Lambda_{\chi}$, $m_{l}/\Lambda_{\chi}$, and $\Lambda_{\text{QCD}}/m_{\text{Q}}$, where p is the characteristic momentum of the interaction, $m_{{l}}$ is the mass of the light Goldstone bosons, $m_{\text{Q}}$ is the heavy quark mass, $\Lambda_{\chi}$ is the chiral symmetry breaking scale, and $\Lambda_{\text{QCD}}$ is the characteristic scale of QCD. In a finite volume this expansion scheme is consistent in the p-regime \cite{ Beane, beane2}. 

However, for volumes smaller than the Compton wavelength of the Goldstone bosons, the zero momentum mode is enhanced with respect to the non-zero modes, and an alternative expansion scheme must be utilized %in the chiral limit%
 \cite{gasser}. The regime where the pion zero-modes must be integrated over explicitly while still treating the non-zero modes perturbatively is known as the $\epsilon$-regime \cite{gasser, hansen1,hansen2, hansen3,chipt3,leut1,2+1.1,2+1.2,2+1.3}.  In the $\epsilon$-regime, a new expansion parameter is introduced, $\epsilon\sim 2\pi/\text{L}\Lambda_{\chi}\sim 2\pi/\beta\Lambda_{\chi}$ and $\epsilon^2\sim m_{l}/\Lambda_{\chi}$, where L and  $\beta$ are the spatial and temporal extents, respectively. 
At leading order, one may associate $\Delta_{\text{Q}}$, the hyperfine splitting between the pseudoscalar meson ${P}$ ($\{D^0, D^+, D^+_s\}$, $\{{B}^-,\bar{B}^0,\bar{B}^0_s\}$) and its respective vector meson ${P}^*$ ($\{D^{*0},{D^{*+}},D^{*+}_s\}$, $\{{B}^{*-}, \bar{B}^{*0},\bar{B}^{*0}_s\}$), with the physical values on the order of 140 MeV and 50 MeV for the charm and bottom mesons respectively.  Therefore, it is reasonable to expect the hyperfine splitting to contribute at order $\epsilon^2$ for charmed mesons ($\epsilon^2 \sim\Delta_{\text{c}}/\Lambda_{\chi}$) and approximately at order $\epsilon^3$ for bottom mesons ($\epsilon^3 \sim\Delta_{\text{b}}/\Lambda_{\chi}$). For the sake of generality, both scenarios are considered.
 
This study presents the volume dependence of the heavy meson masses at next-to-leading order (NLO), $\mathcal{O}(\epsilon^4)$, in the ``mixed regime" of  \text{SU}(2) and \text{SU}(3) HM$\chi$PT. In the ``mixed regime", the physical pion mass is small compared to the IR cutoff and therefore fall within the $\epsilon$-regime, while the  kaon and eta still satisfy the p-regime criteria \cite{mixedregime1, mixedregime2}. Therefore in the mixed regime, the expansion in the $\pi$ and $\{K, \eta\}$ masses is treated separately in order to satisfy $\epsilon^2 \sim m_{\pi}/\Lambda_{\chi} $ and $\epsilon\sim m_K/\Lambda_{\chi}\sim m_{\eta}/\Lambda_{\chi}$. The $\mathcal{O}(\epsilon^3)$ volume dependence of the heavy meson mass for an $\text{SU}(2)_{L}\times \text{SU}(2)_{R}$ chiral theory with  static heavy quarks has been previously calculated \cite{mixedregime2}. Unfortunately, the $\mathcal{O}(\epsilon^3)$ result does not give the predictive power necessary to extract the LECs desired.

%%%%%%%%%%%%%%%%%%%%%%%%%%%%%%%%%%%%%%%%%%%%%%%%%%
\vspace{-.5cm}
\section{Heavy Meson Chiral Perturbation Theory }
The  field multiplet of the pseudoscalar ${P}$ and the vector ${P}^*$  
can be conveniently represented as a single field operator \cite{heavychiPT1, heavychiPT2, heavychiPT3},
\begin{eqnarray}
{H_a}\equiv \frac{1+\not \hspace{-.05cm} v}{2}
\left[ \not \hspace{-.1cm} {P}_a ^*+i {P}_a \gamma_5 \right],\hspace{.5cm}
{\bar{H}_a}\equiv \gamma^0 H^{\dag}_a\gamma^0,
%=\left[  \not \hspace{-.1cm} {P}^{*\dag}_a +i {{P}^{\dag}_a} \gamma_5 \right]\frac{1+\not \hspace{-.05cm} v}{2}
  \label{H_field}
\end{eqnarray}  
where $v^{\mu}$ is the velocity of the heavy meson.
The representation of the composite field $H_a(x)$ assures it transforms as an \text{SU}(2) spinor under heavy-quark spin rotations, under the unbroken $\text{SU}(3)_V$ symmetry it transforms as an element of the  $\bar{\textbf{3}}$ fundamental representation (as denoted by the subscript ``$a$"), and under both Lorentz and parity transformations it is a bilinear. %In adopting the Dirac basis for the gamma matricesand when considering the rest frame, $v^{\mu}=(1,0)$, one arrives at:\begin{eqnarray}H=\begin{pmatrix}0 & i{P}-\vec{\sigma} \cdot\vec{{P}}^* \\0 & 0 \\\end{pmatrix} ,\hspace{1cm}\bar{H}=\begin{pmatrix}0 & 0 \\i{P}^{\dag}+\vec{\sigma} \cdot\vec{{P}}^{*\dag}& 0 \\\end{pmatrix}.\label{fieldsrest}\end{eqnarray} 
%This result is the same for both Minkowski and Euclidian space. 
In the rest frame, the LO HM$\chi$PT Lagrangian in $m_{\text{Q}}$ and $\Lambda_{ \chi}$  consistent with spontaneously broken $\text{SU}(3)_{L}  \times \text{SU}(3)_{R}$ is \cite{heavychiPT1, heavychiPT2, heavychiPT3}:
\begin{eqnarray}
%  {\mathcal L}^0_E= -Tr\left[ \bar{H}_a \partial^0 H_a\right]+{g} \hspace{.1cm} Tr\left[ \bar{H}_a  H_b \vec{\gamma}\cdot\vec{\mathcal{A}} _{ba} \gamma_5\right]  +\frac{f^2}{8}Tr\left[\partial^\mu \Sigma^{\dag}\partial^\mu\Sigma\right] -\frac{f^2}{4}Tr\left[ \mathcal{M}\Sigma^{\dag}+h.c\right],  
  {\mathcal L}^0= 
  -iTr\left[ \bar{H}_a (D^0)_{ba} H_b\right]
  -{g} \hspace{.1cm} Tr\left[ \bar{H}_a  H_b \vec{\gamma}\cdot\vec{\mathcal{A}} _{ba} \gamma_5\right]
  +\frac{f^2}{8}Tr\left[\partial^\mu \Sigma^{\dag}\partial_\mu\Sigma\right]
    +\frac{f^2}{4}Tr\left[ \mathcal{M}\Sigma^{\dag}+h.c\right],      
    \label{lagrangian_0_e}
      \end{eqnarray}     
where $\mathcal{M}= \frac{1}{2}diag(m_{\pi}^2,m_{\pi}^2, 2m_{K}^2 -m_{\pi}^2)$ is the light meson mass matrix, $\vec{\gamma}$ is the spatial component of $\gamma^{\mu}$, $D^{\mu}=\partial^{\mu}+\mathcal{V}^{\mu}$ is the covariant derivative, $f$ is the pion decay constant, and the Goldstone bosons are encapsulated in the operators, 
\begin{eqnarray}
{\Sigma}={\xi}^2=\exp\left(\frac{2i{M}}{f}\right) &\hspace{1cm}&
{M}=
\begin{pmatrix} 
\frac{{\pi}^0}{\sqrt{2}} +\frac{{\eta}}{\sqrt{6}} &{\pi}^+ &{K}^+\\
{\pi}^- & -\frac{{\pi}^0}{\sqrt{2}}+\frac{{\eta}}{\sqrt{6}}&{K}^0 \\
{K}^- &{\bar{K}}^0&-\sqrt{\frac{2}{3}}{\eta} \\
\end{pmatrix} \nonumber\\
{\mathcal{A}}^{\mu}=\frac{i}{2}\left({\xi}\partial^{\mu}{\xi}^{\dag}-{\xi}^{\dag}\partial^{\mu}{\xi}\right)
&\hspace{1cm}&
{\mathcal{V}}^{\mu}=\frac{1}{2}\left({\xi}\partial^{\mu}{\xi}^{\dag}+{\xi}^{\dag}\partial^{\mu}{\xi}\right).
 \end{eqnarray} 
 At NLO in HM$\chi$PT there are a large number of corrections to the Lagrangian that are consistent with velocity reparametrization invariance (VRI) \cite{lagrangian1}, but the terms that will contribute to the volume dependance of the mass are the following:
\begin{eqnarray}
  {\mathcal L}^1&=& 
 - \frac{g_1}{m_{\text{Q}}} \hspace{.1cm} Tr\left[ \bar{H}_a  H_b \vec{\gamma}\cdot\vec{\mathcal{A}} _{ba} \gamma_5\right]
- \frac{ g_2}{m_{\text{Q}}}  Tr\left[ \bar{H}_a  \vec{\gamma}\cdot\vec{\mathcal{A}}  _{ba} \gamma_5 H_b \right]
 + \frac{ \lambda}{m_{\text{Q}}}  Tr\left[ \bar{H}_a \sigma^{\mu\nu}H_a \sigma_{\mu\nu}\right]
 +\frac{\gamma_1}{\Lambda_{\chi}}Tr\left[ \bar{H}_a  H_a\right] \left(\mathcal{A} ^0\mathcal{A}^0\right)_{bb}
 \label{lagrangian_1_e}
\\
 &+&\frac{\sigma_1}{{\Lambda_\chi}} Tr\left[ \bar{H}_a H_b \left(\xi\mathcal{M}\xi+h.c.\right)_{ba} \right]
 + \frac{\sigma_2}{\Lambda_\chi}Tr\left[ \bar{H}_a H_a \left(\xi\mathcal{M}\xi+h.c.\right)_{bb}\right]
  +\frac{\gamma_2}{\Lambda_{\chi}}Tr\left[ \bar{H}_a  H_c \mathcal{A} _{cb}^0\mathcal{A} _{ba}^0 \right]
   +\frac{\gamma_3}{\Lambda_{\chi}}Tr\left[ \bar{H}_a  H_c  \mathcal{A} _{cb}\cdot\mathcal{A} _{ba} \right]\nn\\
    &+&\frac{\gamma_4}{\Lambda_{\chi}}Tr\left[ \bar{H}_a  H_a\right] \left(\mathcal{A} \cdot\mathcal{A}\right)_{bb}\nn.
\end{eqnarray} 
At leading order, the hyperfine splitting can be written in terms of the LEC $\lambda$, $\Delta_{\text{Q}}\equiv \frac{8\lambda}{m_{\text{Q}}}$.

%%%%%%%%%%%%%%%%%
\section{Zero-Modes Integration in the $\epsilon$-regime}
In the $\epsilon$-regime, it is necessary to evaluate the pion zero-modes, $q^{\mu}=(0, \vec{0})$, contribution non-perturbatively. It is convenient  to integrate zero-mode out of the theory, leaving an effective field theory in terms of the non-zero modes. In the mixed regime, only the pion zero-modes are removed \cite{mixedregime2}, while the zero-modes of the kaon and eta are treated perturbatively. This can be done by rewriting the $\Sigma$ operator as
\begin{eqnarray}
\Sigma(x)=U\hat{\Sigma}(x)U,\hspace{1cm}
U=\exp\left[\frac{i}{f}
\begin{pmatrix} 
\frac{{\pi}_z^0}{\sqrt{2}}  &{\pi}_z^+ &0\\
{\pi}_z^- & -\frac{{\pi}_z^0}{\sqrt{2}}&0 \\
0&0&0 \\
\end{pmatrix}\right].
\label{redef1}
 \end{eqnarray} 
The subscript z denotes zero-mode operators, while the operators with a hat are operators whose contribution can be treated perturbatively in the $\epsilon$-expansion. When integrating over the zero-modes it is convenient to write the operator U in terms of hyperspherical coordinates, 
 \begin{eqnarray}
 U&=&
\begin{pmatrix} 
 \cos(\psi)+{i}\cos(\theta)\sin(\psi)&\sin(\theta)\sin(\psi)e^{i\phi}&0\\
\sin(\theta)\sin(\psi)e^{-i\phi}& \cos(\psi)-{i}\cos(\theta)\sin(\psi)&0 \\
0&0&1 \\
\end{pmatrix}.
 \end{eqnarray}  
When constructing the Lagrangian that is invariant under chiral transformations, it is advantageous to define the operator $\xi\equiv\sqrt{\Sigma}$. Under the redefinition of Eq. (\ref{redef1}), one finds, 
 \begin{eqnarray}
 \label{redef2}
 \xi(x)=U\hat{\xi}(x)V^{\dag}(x)=V(x)\hat{\xi}(x)U,
 \end{eqnarray}
 where the definitions $V^{\dag}=\hat{\xi}U^{\dag}\sqrt{U\hat{\Sigma}U}$ and  $V=\sqrt{U\hat{\Sigma}U}\;U^{\dag}\hat{\xi}^{\dag}$ have been implicitly introduced. When integrating over $U$, one may substitute  $\mathcal{A}_{\mu}=V\hat{\mathcal{A}}_{\mu}V^{\dag}$ and $\mathcal{V}_{\mu}=V\hat{\mathcal{V}}_{\mu}V^{\dag}+iV\partial_{\mu}V^{\dag}$.
The results presented here will be truncated at $\mathcal{O}(\epsilon^4)$, in which case one can safely make the following approximation:
\begin{eqnarray} 
\mathcal{A}^{\mu} \simeq\hat{\mathcal{A}}^{\mu}=\frac{i}{2}\left(\hat{\xi}\partial^{\mu}\hat{\xi}^{\dag}-\hat{\xi}^{\dag}\partial^{\mu}\hat{\xi}\right)= \frac{\partial^{\mu}\hat{M}}{f}+\mathcal{O}(\epsilon^3)
&\hspace{1cm}&
 \mathcal{V}^{\mu}\simeq\hat{\mathcal{V}}^{\mu}=\frac{1}{2}\left(\hat{\xi}\partial^{\mu}\hat{\xi}^{\dag}+\hat{\xi}^{\dag}\partial^{\mu}\hat{\xi}\right)= \frac{\hat{M}\partial^{\mu}\hat{M}}{f^2}+\mathcal{O}(\epsilon^4).
 \end{eqnarray} 
The only contribution to the heavy meson mass that originates from the zero-modes integration appears at $\mathcal{O}(\epsilon^4)$, and comes from the second line in Eq. (\ref{lagrangian_1_e}):
 \begin{eqnarray}
 \label{splitting}
\delta\mathcal{L}_\mathcal{M}&=&
  \frac{\sigma_1}{\Lambda_{\chi}}  Tr\left[ \bar{H}_a H_b \left(\xi\mathcal{M}\xi+h.c.\right)_{ba} \right]
 + \frac{\sigma_2}{\Lambda_{\chi}}Tr\left[ \bar{H}_a H_a \left(\xi\mathcal{M}\xi+h.c.\right)_{bb}\right] \nonumber\\
  &\simeq&
\frac{\sigma_1}{\Lambda_{\chi}}Tr[\bar{H}_1H_1+ \bar{H}_2H_2]\left(\cos(2\psi)m_{\pi}^2\right)
+\frac{\sigma_1}{\Lambda_{\chi}}Tr[\bar{H}_3H_3]\left(2m_{k}^2-m_{\pi}^2\right)\\
&&+\frac{\sigma_2}{\Lambda_{\chi}}Tr[\bar{H}_aH_a]\left(2m_K^2+\left(-1+2\cos(2\psi)\right)m_{\pi}^2\right)\nn
 .
\end{eqnarray}
In order to evaluate the contribution of this term to the heavy meson mass, the $\psi$-dependence in this expression must be integrated out using the non-perturbative weight arising from the last term in Eq. (\ref{lagrangian_0_e}). It is convenient to perform this integral by analytically continuing to Euclidean time $t\rightarrow -it$ :
\begin{eqnarray}
 && \int \mathcal{D}U^2\; \exp\left[\int
  d^4x\left(\frac{f^2 }{4}Tr\left[\mathcal{M}U^2+h.c\right] +\delta\mathcal{L}_\mathcal{M}\right)\right]\nonumber\\
&=& X(s)\; \exp\left[
-\frac{X^\prime(s)}{X(s)} m_{\pi}^2 \int d^4x
  \left(  \left( \sigma_1+2\sigma_2\right) \left({P}^{*\dag}_a{P}^{*}_a +\vec{{P}}^{\dag}_a\cdot\vec{{P}}_a\right)
- \sigma_1\left({P}^{*\dag}_3{P}^{*}_3 +\vec{{P}}^{\dag}_3\cdot\vec{{P}}_3\right)\right)
\right].
\end{eqnarray}
where $s=\frac{1}{4}{f^2m_{\pi}^2\beta \text{L}^3}$, and X(s) can be expressed in terms of the modified Bessel function $I_1(2s)$ of the first kind,
 \begin{eqnarray}
X(s)\equiv
\int \frac{8}{\pi} d\psi \cos^2(\psi)\sin^2(\psi)\; e^{2s\cos(2\psi)}=\frac{I_1(2s)}{s},
%=\frac{I_1(2s)}{2s}
&\hspace{.2cm}&
\int \frac{8}{\pi} d\psi \cos^2(\psi)\sin^2(\psi)\; e^{2s\cos(2\psi)}\cos(2\psi)=\frac{X'(s)}{2}.
\end{eqnarray}
%Only the finite-volume dependence has been written.  
Since $\mathcal{M}$ has no spin structure, it results in the same shift for both the masses of the pseudoscalar and vector fields,  yet it explicitly breaks the $\text{SU}(3)_V$ symmetry,  
\begin{eqnarray}
\delta \text{M}_{({P}, {P}^*)}= \frac{m_{\pi}^2}{\Lambda_{\chi}}  (\sigma_1+2\sigma_2)\frac{X^\prime(s)}{X(s)} 
+\frac{\sigma_2}{\Lambda_{\chi}}2(2m_K^2-m_{\pi}^2),
\hspace{.4cm}
\delta \text{M}_{({P_{\bar{s}}}, P_{\bar{s}}^*)}= 2\frac{m_{\pi}^2}{\Lambda_{\chi}} \sigma_2\frac{X^\prime(s)}{X(s)}
+\frac{\sigma_1+\sigma_2}{\Lambda_{\chi}}2(2m_K^2-m_{\pi}^2)
\label{zero_modes}.
\end{eqnarray}
This analysis introduces a volume dependence to the mass of the non-zero pion modes, as well as for the $K$'s and $\eta$'s,
\begin{eqnarray} 
\label{effectivepionmass}
m_{\pi}^2\rightarrow m_{\pi}^2\frac{X^\prime(s)}{2X(s)}, \hspace{1cm} 
m_{K}^2&\rightarrow& m_{K}^2 -\frac{m_{\pi}^2}{2}+m_{\pi}^2\frac{X^\prime(s)}{4X(s)}= m_{K}^2+\mathcal{O}(\epsilon^4) \\
\label{effectiveetamass}m_{\eta}^2 &\rightarrow& \frac{4}{3}m_{K}^2 -2\frac{m_{\pi}^2}{3}+m_{\pi}^2\frac{X^\prime(s)}{6X(s)}
= \frac{4}{3}m_{K}^2+\mathcal{O}(\epsilon^4).
\end{eqnarray}
 After performing the integration over the zero-modes, the finite-volume contribution from the remaining degrees of freedom can be evaluated perturbatively. The finite volume Feynman diagrams can be evaluated in the standard way, where the integral is replaced by a sum over discretized four momenta and the zero mode is explicitly excluded in the pion loops \cite{Beane}. An outline of the methods used in performing these sums is discussed in the appendix. Eq. (\ref{zero_modes}) also includes a volume independent $\text{SU}(3)_V$ symmetry breaking term, 
 \begin{eqnarray}
 \label{pseudo_strange}
 \delta^s_Q\equiv
( \delta \text{M}_{P_{\bar{s}}}- \delta \text{M}_{P})_{(L= \infty)}= {2\sigma_1}\frac{(2m_K^2-m_{\pi}^2)}{\Lambda_{\chi}}.
 \end{eqnarray} 
 This results in a shift on the bare mass  of the strange pseudoscalar meson, $\delta^s_Q$, and the strange vector meson 
 \begin{eqnarray}
  \label{strange_vector}
 \Delta^{s}_{\text Q}\equiv\Delta_{\text{Q}}+\delta^s_Q.
  \end{eqnarray}
 At leading order, one may associate $\delta ^{s}_Q$ with physical value of the splitting between the isospin doublet ${P}$ and strange pseudoscalar  $P_{\bar{s}}$, which is on the order of 100 MeV for both the charm and bottom mesons respectively.  For the sake of generality, both $\delta ^{s}_Q$ and $\Delta ^{s}_Q$ will assume the same power counting as $\Delta_{\text{Q}}\sim\mathcal{O}(\epsilon^2)$.

%%%%%%%%%%%%%%%%%
%%%%%%%%%%%%%%%%%
\begin{figure}[htp]
\centering
\includegraphics[totalheight=1.2cm]{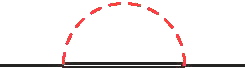}
\caption{$\epsilon^3$ contribution to the pseudoscalar heavy meson mass. The solid line corresponds to the heavy pseudoscalar, the double line denotes a vector meson, and the dashed line represents a Goldstone boson.}\label{leading}
\end{figure}
\section{Results} 
In taking the isospin limit, the pseudoscalar pair $P=\{P_{\bar{u}}, P_{\bar{d}} \}$ will receive the same mass contribution. $P$ will denote the isospin pair and $P_{\bar{s}}$ will denote the strange-heavy meson. In order to formally categorize the different terms contributing to the mass, it is important to consider the ratio ${\Lambda_{\text{QCD}}}/{m_{\text{Q}}}\sim\mathcal{O}(\epsilon^{\alpha})$. The most relevant cases are the following:
\begin{eqnarray}
\label{cases}
(i):\;\Lambda_{\text{QCD}}/m_{\text{Q}}\sim\mathcal{O}(\epsilon^{2})
\hspace{1cm}
(ii):\;\Lambda_{\text{QCD}}/m_{\text{Q}}\sim\mathcal{O}(\epsilon),
\label{cases}
\end{eqnarray}
corresponding to the static limit and LO heavy quark mass corrections, respectively. For simplicity, the expressions below will include finite LO heavy quark mass corrections. The static limit can easily be obtained by taking the $m_{\text{Q}}\rightarrow \infty$ limit (note $\Delta_{\text{Q}}\propto m_{\text{Q}} ^{-1}\rightarrow 0$). %Due to field conventions, an additional factor of $\frac{1}{2}$ is needed. 
The individual diagrams contributing to $M_ P$ are written in the appendix. The notation $\delta \text{M}_{P}\equiv M_{P}(L)- M_{P}(\infty)$ is used to denote the finite-volume dependence of the mass. 

\subsection{1. \text{SU}(3) HM$\chi$PT}
\begin{figure}[htp]
\centering
\includegraphics[totalheight=2cm]{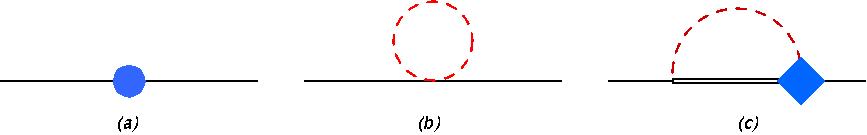}
\caption{$\epsilon^4$ contribution to the pseudoscalar heavy meson mass. (a) Denotes the zero-modes contribution. (b) Goldstone bosons loops originating from four-point vertices. (c) Incorporates operators that contribute the heavy flavor symmetry breaking corrections to the $PP^*\pi$ vertex.}\label{subleading}
\end{figure}

\begin{figure}[htp]\centering\includegraphics[totalheight=4cm]{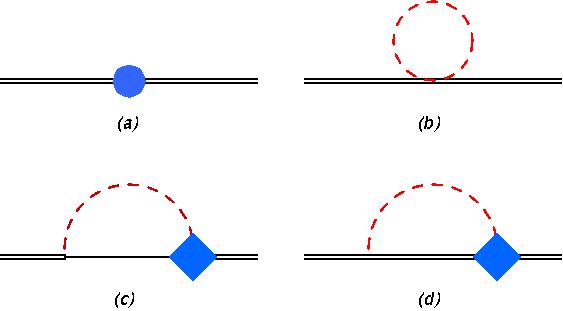}\caption{$\epsilon^4$ contribution to vector meson mass.}\label{vector_subleading} \end{figure}

The \text{SU}(3) volume dependence of the $P$ and $P_{\bar{s}}$ masses up to and including $\mathcal{O}(\epsilon^4)$ is found by adding the finite-volume contributions from the self-energy diagrams depicted in Figs. (\ref{leading}) and (\ref{subleading}), where the Goldstone bosons can be pions, kaons and etas,  
 \vspace{-1cm}
\begin{eqnarray}
\label{results1}
\delta \text{M}_{P}
&=& \left({g^2+2\frac{g(g_1-g_2)}{m_{\text{Q}}}}\right)\left(\frac{3}{4f^2 \text{L}^3}
+\frac{1}{8\pi Lf^2}\left({m_K^2} \mathcal{N}_1(m_K, L)+{\frac{1}{6}m_{\eta}^2} \mathcal{N}_1(m_{\eta}, L)\right)\right)
+\frac{g^2{{2\sigma_1}m_K^2}}{{f^2 \Lambda_{\chi}}}\left(2\frac{\mathcal{N}_2(m_K, L)} {m_K^{1/2} {L^{5/2}}}\right)
\\
&+&
\frac{g^2{\Delta_{\text{Q}}}}{{2f^2 }}\left(-\frac{3c_1}{8\pi L^2}+\frac{1}{3}\frac{\mathcal{N}_2(m_{\eta}, L)} {m_{\eta}^{1/2}{L^{5/2}} }+2\frac{\mathcal{N}_2(m_K, L)} {m_K^{1/2} {L^{5/2}}}\right)
-\frac{\gamma_1}{2f^2\Lambda_{\chi}}\left(6\frac{c_4}{2\pi^2 \text{L}^4}+8 \frac{m_K^3}{16\pi^2 L}\mathcal{K}_1(m_KL)
+2 \frac{m_{\eta}^3}{16\pi^2 L}\mathcal{K}_1(m_{\eta}L)\right)
\nn\\ 
&-&
\frac{\gamma_2}{2f^2\Lambda_{\chi}}
\left(3\frac{c_4}{2\pi^2 \text{L}^4}+2 \frac{m_K^3}{16\pi^2 L}\mathcal{K}_1(m_KL)
+\frac{1}{3} \frac{m_{\eta}^3}{16\pi^2 L}\mathcal{K}_1(m_{\eta}L)\right)
-\frac{\gamma_3}{2f^2\Lambda_{\chi}}\left(
2\frac{m_K^3}{16\pi^2 L}\mathcal{K}_2(m_KL)
+\frac{1}{3}\frac{m_{\eta}^3}{16\pi^2 L}\mathcal{K}_2(m_{\eta}L)
\right) \nn\\
&-&\frac{\gamma_4}{2f^2\Lambda_{\chi}}\left(8 \frac{m_K^3}{16\pi^2 L}\mathcal{K}_2(m_KL)
+2 \frac{m_{\eta}^3}{16\pi^2 L}\mathcal{K}_2(m_{\eta}L)\right)
+ \frac{m_{\pi}^2}{2\Lambda_{\chi}}  (\sigma_1+2\sigma_2)\frac{X^\prime(s)}{X(s)}
,\nn\\
\delta \text{M}_{P_{\bar{s}}}&=&\frac{2}{f^2}\left({g^2+2\frac{g(g_1-g_2)}{m_{\text{Q}}}}\right)\left(
\frac{m_K^2}{8\pi L}\mathcal{N}_1(m_K, L)+
\frac{1}{3}\frac{m_{\eta}^2}{8\pi L}\mathcal{N}_1(m_{\eta}, L)
\right)-
\frac{8g^2 {{\sigma_1}m_K^2}}{f^2 L^{5/2}{\Lambda_{\chi}}}
\frac{ \mathcal{N}_2(m_K, L)}{m_K^{1/2}}
\label{results2}\\
&+&
\frac{2g^2 {\Delta_{\text{Q}}}}{f^2 L^{5/2}}
\left(\frac{\mathcal{N}_2(m_{\eta}, L)
}{3m_{\eta}^{1/2}}
+
\frac{ \mathcal{N}_2(m_K, L)}{m_K^{1/2}}
\right)
-\frac{\gamma_1}{2f^2\Lambda_{\chi}}\left(3\frac{c_4}{2\pi^2 \text{L}^4}+8 \frac{m_K^3}{16\pi^2 L}\mathcal{K}_1(m_KL)
+2 \frac{m_{\eta}^3}{16\pi^2 L}\mathcal{K}_1(m_{\eta}L)\right)
\nn\\ 
&-&\frac{\gamma_2}{2f^2\Lambda_{\chi}}
\left( 4\frac{m_K^3}{16\pi^2 L}\mathcal{K}_1(m_KL)
+\frac{4}{3} \frac{m_{\eta}^3}{16\pi^2 L}\mathcal{K}_1(m_{\eta}L)\right)
-\frac{\gamma_3}{2f^2\Lambda_{\chi}}\left(
4\frac{m_K^3}{16\pi^2 L}\mathcal{K}_2(m_KL)
+\frac{4}{3}\frac{m_{\eta}^3}{16\pi^2 L}\mathcal{K}_2(m_{\eta}L)\right)\nn\\
&-&\frac{\gamma_4}{2f^2\Lambda_{\chi}}\left(8 \frac{m_K^3}{16\pi^2 L}\mathcal{K}_2(m_KL)
+2 \frac{m_{\eta}^3}{16\pi^2 L}\mathcal{K}_2(m_{\eta}L)\right)
+2\sigma_2  \frac{m_{\pi}^2}{2\Lambda_{\chi}}  \frac{X^\prime(s)}{X(s)}.\nn
 \end{eqnarray}
where $m_{\eta}^2\equiv4m_K^2/3$, the discrete sums $c_1$, $c_4$, $\mathcal{N}_i$, and $\mathcal{K}_i$ are defined in Eq. (\ref{sums}), (\ref{nfnctions}-\ref{kfnctions}), and Eq. (\ref{pseudo_strange}-\ref{strange_vector}) have been used. 
%%%%%%%%%%%%%%%%%%%%%%%%%%%%%%%%%%%%%%%%%%%%%%%%%%%
\begin{figure}[htp]
\centering
\includegraphics[totalheight=2cm]{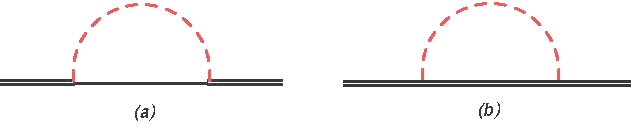}
\caption{$\epsilon^3$ contribution to the heavy vector meson mass.}\label{vector_Leading}
\end{figure} 
Similarly, the $\mathcal{O}(\epsilon^3)$ and $\mathcal{O}(\epsilon^4)$ corrections to the vector meson masses are depicted in Fig. (\ref{vector_Leading}) and Fig. (\ref{vector_subleading}), respectively. 
In the static limit pseudoscalar and the vector mesons are degenerate, therefore it is only necessary to evaluate the volume dependence of the hyperfine splitting:
\begin{eqnarray}
\label{results3}
\delta \text{M}_{P^*}-\delta \text{M}_{P}&=&
{8\frac{gg_2}{3m_{\text{Q}}}}\left(\frac{3}{4f^2 \text{L}^3}
+\frac{1}{8\pi Lf^2}\left({m_K^2} \mathcal{N}_1(m_K, L)+{\frac{1}{6}m_{\eta}^2} \mathcal{N}_1(m_{\eta}, L)\right)\right)
\\\nn
&+&\frac{g^2{\Delta_{\text{Q}}}}{{2f^2 }}\left(\frac{c_1}{2\pi L^2}
-\frac{4}{3}\frac{\mathcal{N}_2(m_K, L)} {m_K^{1/2} {L^{5/2}}}+\frac{2}{9}\frac{\mathcal{N}_2(m_{\eta}, L)} {m_{\eta}^{1/2}{L^{5/2}} }\right)
\\ 
\label{results4}
\delta \text{M}_{P_{\bar{s}}^*}-\delta \text{M}_{P_{\bar{s}}}&=&\frac{16gg_2}{3m_{\text{Q}}f^2}\left(
\frac{m_K^2}{8\pi L}\mathcal{N}_1(m_K, L)+
\frac{1}{3}\frac{m_{\eta}^2}{8\pi L}\mathcal{N}_1(m_{\eta}, L)
\right)
\\\nn
&-&\frac{4}{3}
\frac{g^2 {\Delta_{\text{Q}}}}{f^2 L^{5/2}}
\left(2\frac{ \mathcal{N}_2(m_K, L)}{m_K^{1/2}}
+\frac{2}{3}\frac{2g^2 {\Delta_{\text{Q}}}}{3f^2 L^{5/2}}
\frac{\mathcal{N}_2(m_{\eta}, L)
}{m_{\eta}^{1/2}}\right)
.  \end{eqnarray}
 
 \subsection{2. \text{SU}(2) HM$\chi$PT}
In \text{SU}(2) chiral perturbation theory, the kaons and eta decouple from the theory. Only integrals including pions depicted in Figs. (\ref{leading}), (\ref{subleading}), (\ref{vector_Leading}), and (\ref{vector_subleading})  contribute to the volume dependence of the masses, 
\begin{eqnarray}\delta \text{M}_{P}^{(2)}
\label{results5}
&=& \left({(g^{(2)})^2+2\frac{g^{(2)}(g^{(2)}_1-g^{(2)}_2)}{m_{\text{Q}}}}\right)\frac{3}{4f^2 \text{L}^3}+
 \frac{m_{\pi}^2}{2\Lambda_{\chi}}  (\sigma^{(2)}_1+2\sigma^{(2)}_2)\frac{X^\prime(s)}{X(s)}\nn
 -\frac{(g^{(2)})^2{\Delta^{(2)}_Q}}{{2f^2 }}\frac{3c_1}{8\pi L^2}
 \\
&&
-\frac{3}{2f^2\Lambda_{\chi}}\left(2\gamma^{(2)}_1+\gamma^{(2)}_2\right)\frac{c_4}{2\pi^2 \text{L}^4},
\\\label{results6}
\delta \text{M}_{P^*}^{(2)}-\delta \text{M}_{P}^{(2)}&=&
{2\frac{g^{(2)}g^{(2)}_2}{m_{\text{Q}} f^2 \text{L}^3}}+\frac{(g^{(2)})^2{\Delta^{(2)}_Q}}{{f^2 }}\frac{c_1}{4\pi L^2},
\end{eqnarray}
where an additional superscript has been introduced in order to explicitly distinguish the \text{SU}(2) LECs from those contributing to the \text{SU}(3) theory.  

Note, these results have been derived assuming $\Delta_{\text{Q}} \sim\mathcal{O}(\epsilon^2)$, which should be expected to be the case for the charm mesons. For the bottom mesons one should expect $\Delta_{\text{b}} \sim\mathcal{O}(\epsilon^3)$. This would move finite volumes effects related to this coupling to  $\mathcal{O}(\epsilon^5)$, displacing them outside the scope of this calculation.

%\begin{eqnarray}\frac{\left<Q(t)\gamma^i\gamma^5 \bar q(t) \bar Q(0)\gamma^i\gamma^5{q}(0)\right>}{\left<Q(t)\gamma^5 \bar q(t) \bar Q(0)\gamma^5q(0)\right>}{\sim}e^{-t(M_{P_q^*}- M_{P_q})},\hspace{.5cm}\frac{\left<Q(t+1)\gamma^i\gamma^5 \bar q(t+1) \bar Q(0)\gamma^i\gamma^5{q}(0)\right>}{\left<Q(t) \gamma^j\gamma^5 \bar q(t) \bar Q(0) \gamma^j\gamma^5q(0)\right>}{\sim} e^{-M_{P_q^*}}.\nn\end{eqnarray}
 %%%%%%%%%%%%%%%%%%%%%%%%%%%%%%%%%%%%%%%%%%%%%%%%%%%%%%%%%%%%
  
   \section{ANALYSIS and Discussion}

The results presented in the previous section allow determination of LECs that play an important role in heavy-light meson scattering. In order to evaluate the LECs, one must fit the expressions $\text M_{\infty}+\delta \text M(\text{L}, m_{\pi})$ to LQCD results of the heavy meson masses for different volumes and pion masses that fall within the $\epsilon$-regime, where $M_{\infty}$ is the infinite volume mass of the heavy meson and $\delta \text M(\text{L}, m_{\pi})$ is the finite volume contribution described by Eqs. (\ref{results1}-\ref{results6}). 

\begin{figure}
\centering
\subfloat[]{\label{fit_mass}\includegraphics[totalheight=4.4cm]{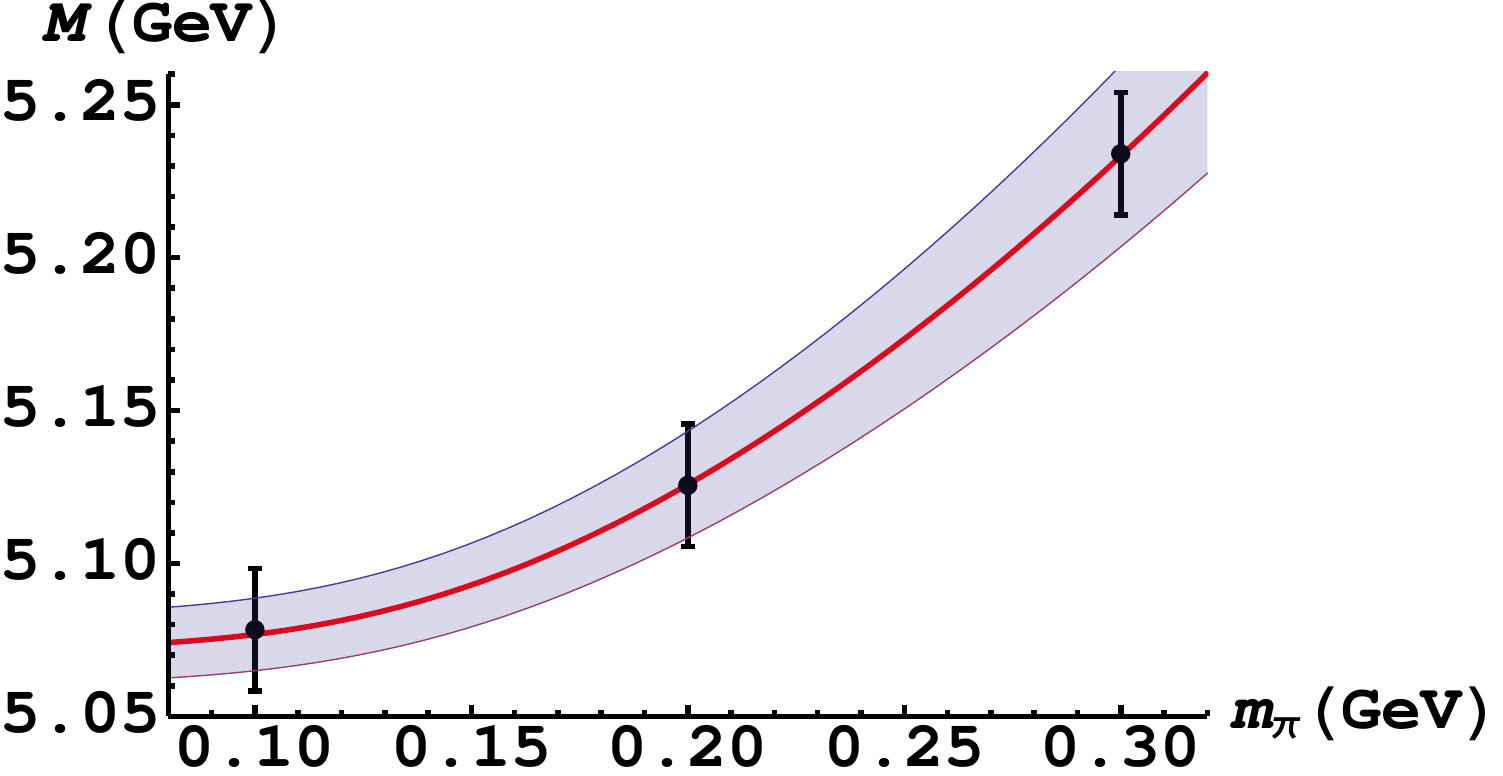}}
\hspace{1cm} 
\centering
\subfloat[]{\label{fit_volume}\includegraphics[totalheight=4.4cm]{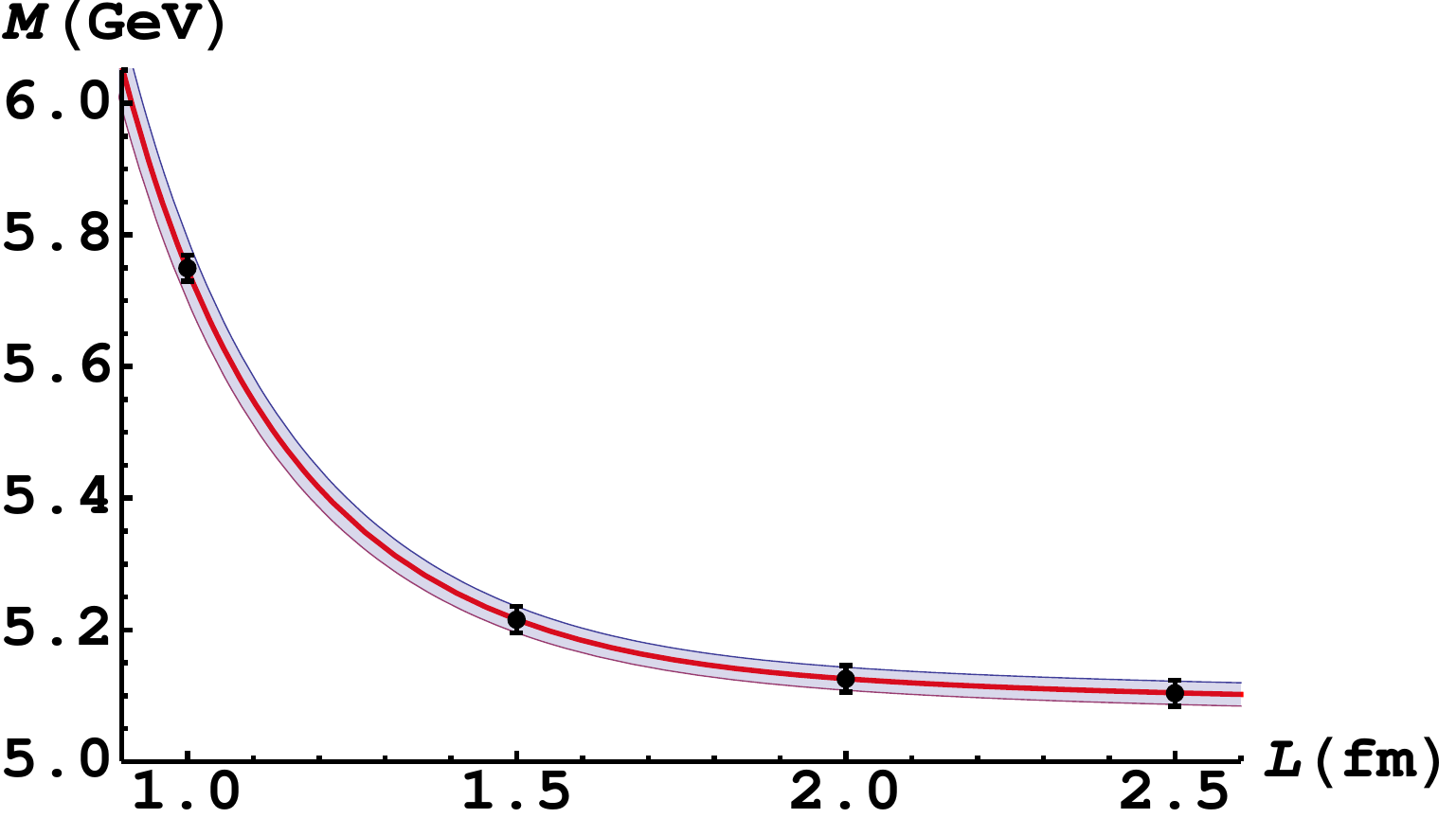}}
\caption[alignment]{	In black is the fake data with corresponding error bars of approximately  $0.5\%$. The red line  is the fit of Eq. (\ref{results7}) to the data, and the filled area denotes the error associated with the fit of the LECs and $\text{M}_{\infty}$.  The volume chosen for the data depicted in Fig. (\ref{fit_mass}) is 2 fm, and the pion mass chosen for the data depicted in Fig. (\ref{fit_volume}) is 0.2 GeV. }
\label{FIT} \end{figure}

\begin{figure}
\centering
\subfloat[]{\label{fit_mass}\includegraphics[totalheight=4.4cm]{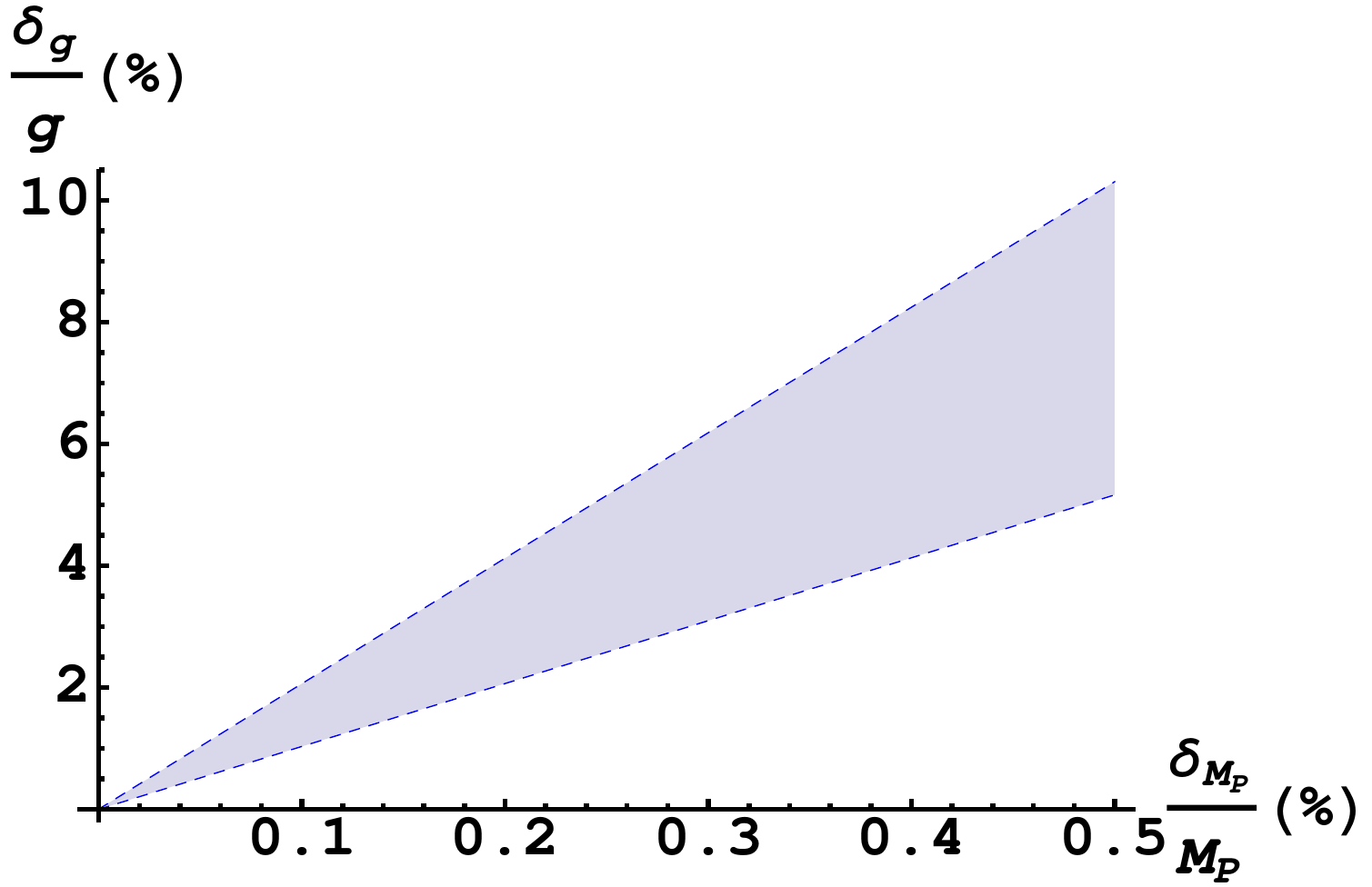}}
\hspace{1cm} 
\centering
\subfloat[]{\label{fit_volume}\includegraphics[totalheight=4.4cm]{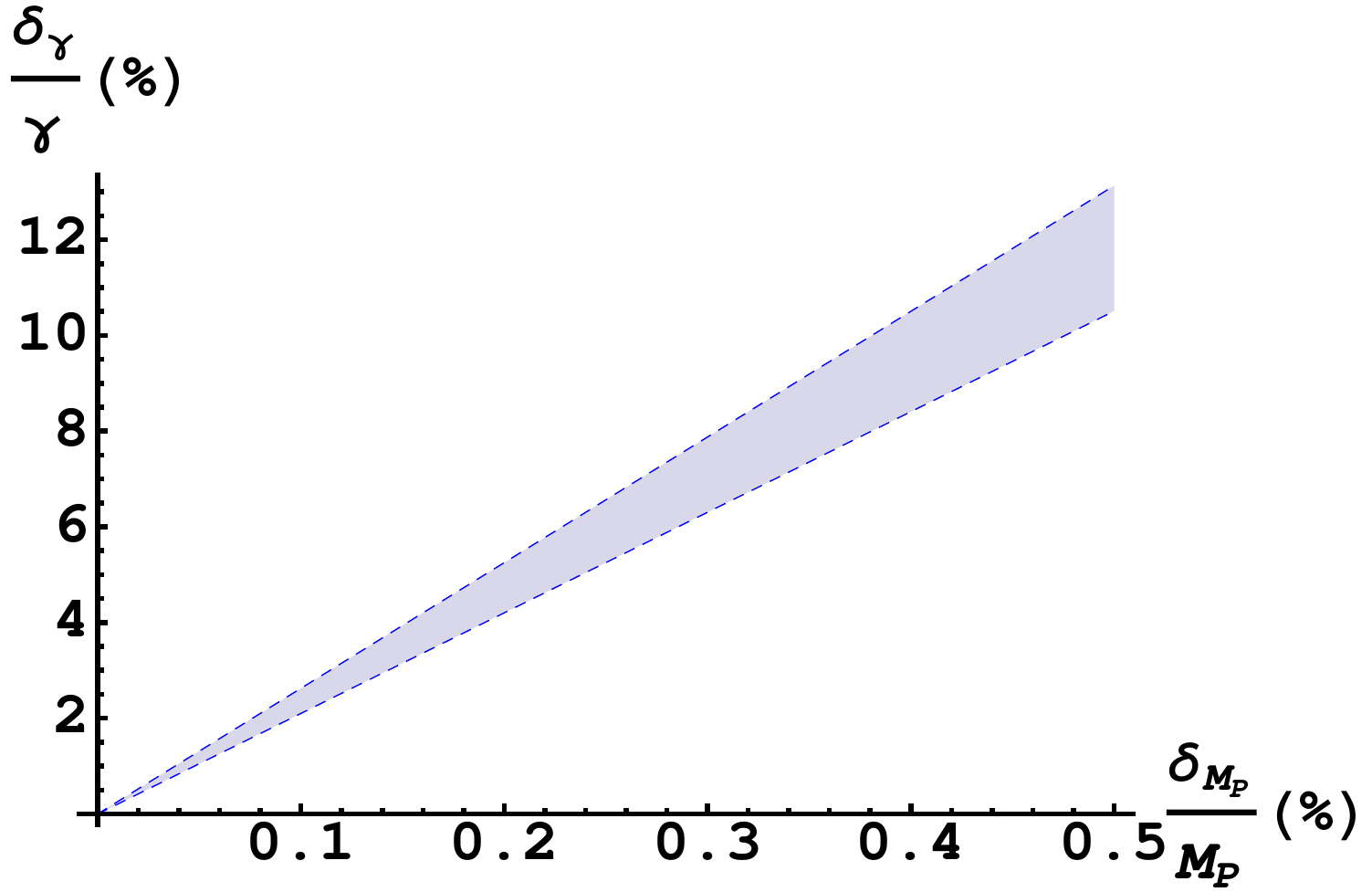}}
\centering
\subfloat[]{\label{fit_volume}\includegraphics[totalheight=4.4cm]{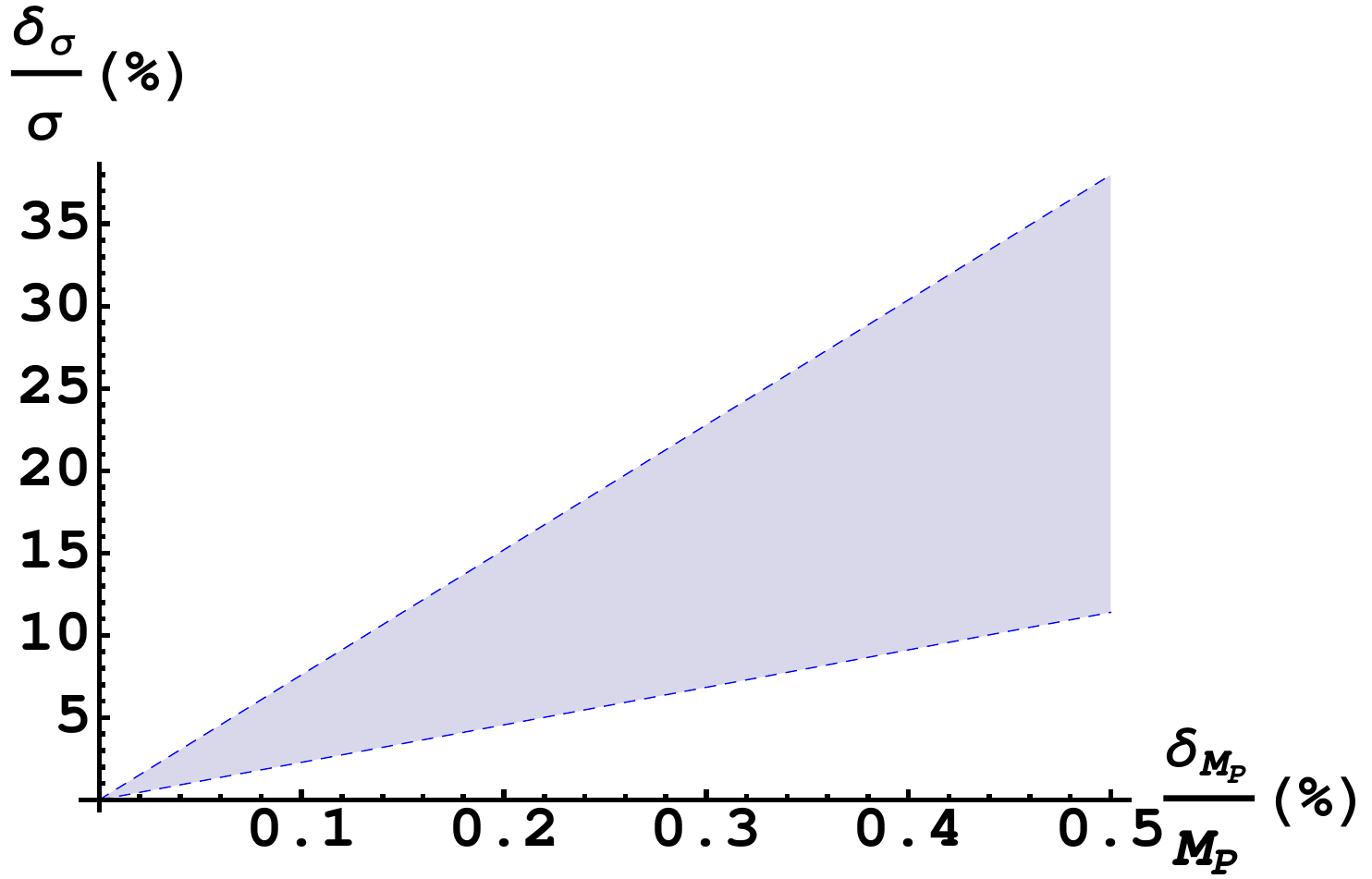}}

\caption[alignment]{Prediction for the level of precision for determining \{g, $\gamma, \sigma\}$ by fitting Eq. (\ref{results7}) to six heavy mesons masses, with an uncertainty ranging from 0.5\% to 0.01\%, and each located at a different (\text{L}, $m_{\pi}$) that fall within the $\epsilon$-regime. The LECs used in generating the fake data were allowed to vary $\text{M}_{\infty}=\{4.5,5.5\}$ {GeV}, ${f}=130$ {MeV}, $\text{g}=\{0.5,1.5\}$, ${\gamma}=\{0.5,1.5\}$ $\text{GeV}^{-1}$, and ${\sigma}=\{0.5,1.5\}$ $\text{GeV}^{-1}$. }
\label{FIT2} \end{figure}

The corresponding LQCD calculation has not been performed yet. Nevertheless, the uncertainty of the LECs as a function of the standard deviation of the heavy meson masses can be estimated. In order to do this I have analyzed fake data for the heavy meson mass. A data set was generated that follows the trend predicted by $\text M_{\infty}+\delta \text M(\text{L}, m_{\pi})$; this required inputing randomly-generated LECs. 
Additional volume and pion mass dependent terms were added to $\text M_{\infty}+\delta \text M(\text{L}, m_{\pi})$ in order to simulate the $\mathcal{O}(\epsilon^5)$ corrections. The exact form of these terms is irrelevant for the discussion at hand. Each heavy meson mass has been given a corresponding uncertainty. Lastly, the set is fit to $\text M_{\infty}+\delta \text M(\text{L}, m_{\pi})$ in order to reproduce the randomly-generated LECs. Due to the larger number of undetermined LECs and the larger expansion parameters for \text{SU}(3) HM$\chi$PT ($\text m_{\text{K}}/\Lambda_{\chi}$, $\text m_{\eta}/\Lambda_{\chi}$), the following discussion will focus on \text{SU}(2) HM$\chi$PT.
%As expected, this analysis leads to drastically different results for \text{SU}(2) HM$\chi$PT and \text{SU}(3) HM$\chi$PT. From the functional forms of equations Eqs. (\ref{results1}-\ref{results6}) it is clear that one would better determine the obtain better statistics for the \text{SU}(2) case with less data points. Additionally, due to rather large expansion parameters for \text{SU}(3) HM$\chi$PT ($\text m_{\text{K}}/\Lambda_{\chi}$, $\text m_{\eta}/\Lambda_{\chi}$), it is natural to expect large $\mathcal{O}(\epsilon^5)$  corrections, resulting in further uncertainties in the corresponding SU(3) LECs. 

Equations (\ref{results5}-\ref{results6}) depends on three parameters: $\text m_{\pi},$ $\text m_{\text Q},$ and $\text L$. In order to determine  ${\sigma}\equiv{(\sigma^{(2)}_1+2\sigma^{(2)}_2)}/{2\Lambda_{\chi}}$ and ${\gamma }\equiv(2\gamma^{(2)}_1+\gamma^{(2)}_2)/{\Lambda_{\chi}} $, which contribute to the NLO calculation of $P\pi$ scattering lengths \cite{scatteringlengths3, scatteringlengths4}, it is convenient to consider the static limit, $ m_{\text Q}\sim\infty$. In this limit the volume dependence of the mass at $\mathcal{O}(\epsilon^4)$ simplifies to,
\begin{eqnarray}\delta \text{M}_{\infty}^{(2)}(\text{L}, m_{\pi})
\label{results7}
&=& (g^{(2)})^2\frac{3}{4{f}^2 \text{L}^3}+
{m_{\pi}^2} {\sigma}\frac{X^\prime(s)}{X(s)}-\frac{3}{2f^2}{\gamma }\frac{c_4}{2\pi^2 \text{L}^4}.
\end{eqnarray}

Since the level of precision with which the LECs can be determined depends on their magnitude, the parameters were varied $\text{M}_{\infty}=\{4.5,5.5\}$ {GeV}, $\text{g}=\{0.5,1.5\}$, ${\gamma}=\{0.5,1.5\}$ $\text{GeV}^{-1}$, and ${\sigma}=\{0.5,1.5\}$ $\text{GeV}^{-1}$, and the pion decay constant was fixed at ${f}=130$ {MeV}. For each value of the parameters, a set of six heavy meson masses was generated. Each heavy meson mass was calculated with different values of the volume and the pion mass that fall within the $\epsilon$-regime, and each heavy meson mass was given an uncertainty ranging from 0.5\% to 0.01\%. Fig. (\ref{FIT}) serves as an example of one of these sets. The set depicted in Fig. (\ref{FIT}) was generated using the center values for the parameters and each heavy meson mass was given an uncertainty of approximately 0.5\%. The estimate of the expected fractional standard deviation of the LECs as a function of the uncertainty of the masses is plotted in Fig. (\ref{FIT2}) as the shaded region. The range of possible standard deviation for the LECs for a given uncertainty of the heavy meson mass manifests the fact that the precision with which these LECs can be determined depend on their absolute value. 
  
\section{CONCLUSION}

HM$\chi$PT  is the EFT for calculating strong-interaction quantities of heavy mesons. Currently, HM$\chi$PT is limited by the poor determination of the LECs of the theory. In particular, the LECs in the lagrangian discussed in this paper, Eqs. (\ref{lagrangian_0_e} \& \ref{lagrangian_1_e}), contribute to the evaluation of scattering lengths, and are currently known to within a factor of three \cite{scatteringlengths3, scatteringlengths4}. 

These LECs can be evaluated from LQCD calculations. One way to achieve this is to extract the LECS from the volume dependance of the heavy mesons masses, since these finite volume effects are parametrized by the LECs of the theory.  With this in mind, the finite-volume dependence of the heavy pseudoscalar and vector meson masses in the $\epsilon$-regime of HM$\chi$PT have been calculated to  $\mathcal{O}(\epsilon^4)$. In the $\epsilon$-regime, LQCD calculations can be performed at the physical point of QCD ($m_{\pi}\approx 140$ {MeV}) if volumes are small ($\text{L}\leq4 \text{fm}$).
 
Lastly, it was shown that with a set of six heavy mesons masses extracted from LQCD with 0.1\% uncertainties, the SU(2) NLO LECs  ${\sigma}\equiv{(\sigma^{(2)}_1+2\sigma^{(2)}_2)}/{2\Lambda_{\chi}}$ and ${\gamma }\equiv(2\gamma^{(2)}_1+\gamma^{(2)}_2)/{\Lambda_{\chi}}$ can be determined within the $10\%$ level of precision. This level of precision shows the power of this method. In order to calculate the heavy-light scattering lengths, it is also necessary to determine the linear combination $2\gamma^{(2)}_4+\gamma^{(2)}_3$ \cite{scatteringlengths3, scatteringlengths4}, which could be determined from the volume dependance of the heavy meson mass at $\mathcal{O}(\epsilon^6)$.

\subsection{Acknowledgement}

The author would like to thank M. Savage, J. Wasem, and B. Smigielski for many useful conversations. In addition, he is endebted to S. Sharpe, H. W. Lin, A. Jamison, D. Bolton, A. Nicholson,  B. Mattern, and J. Vinson for their helpful comments and discussions. 

 \appendix
 % reset counter 
  \section{APPENDIX}
   \setcounter{equation}{0} 
  \renewcommand{\theequation}{A.\arabic{equation}}

Finite-volume Feynman diagrams can be performed by replacing integrals with sums over discretized four momenta \cite{Beane}. The sums contributing to the calculation of the heavy meson mass to  $\mathcal{O}(\epsilon^4)$ are
\begin{eqnarray}
  {\mathbb A} (m_{\text{l}},\Delta,  L,\beta )&=& \frac{1}{\beta \text{L}^3}
  \sum_{n_\mu\neq 0}\frac{1}{i(\no+\omega)-\Delta} 
  \frac{(\nv)^2}{(\no)^2+(\nv)^2+m_{\text{l}}^2},
  \label{sum1}\\
   {\mathbb B} (m_{\text{l}}, L,\beta )&=& \frac{1}{\beta \text{L}^3}
  \sum_{n_\mu\neq 0}
  \frac{(\no)^2}{(\no)^2+(\nv)^2+m_{\text{l}}^2},  \label{sum2}
  \end{eqnarray}
  \begin{eqnarray}
  {\mathbb C}(m_{\text{l}}, L,\beta )&=& \frac{1}{\beta \text{L}^3}
 \sum_{n_\mu\neq 0}
 \frac{(\nv)^2}{(\no)^2+(\nv)^2+m_{\text{l}}^2}.  \label{sum3}\
    \end{eqnarray}
where $\omega$ is the external energy, and $m_{\text{l}}$ denotes the light meson mass ($m_{\pi}, m_{K}, m_{\eta}$). 
In the mixed regime, the $\pi$ and \{K, $\eta$\} loops must be treated separately. Due to the field convention, the corrections to the mass are defined as $\frac{i}{2}\Pi (\omega=0) $+$\frac{i\delta_P}{2}\partial_{\omega}\Pi(\omega=0)$, where $\Pi$ is the sum of the amputated self-energy diagrams, and $\delta_P$ is the bare residual mass of the heavy meson. The superscripts of the terms below denote the order at which they contribute in the $\epsilon$-expansion. The $\mathcal{O}(\epsilon^3)$ correction to the pseudoscalar mass, depicted in Fig. (\ref{leading}), is
\begin{eqnarray} 
\label{delta1}
M^{(3)}(m_{\pi},\Delta, L,\beta)&=&\frac{1}{2}\left(\frac{2\tilde{g}}{f}\right)^2\frac{3}{2}\frac{1}{\beta \text{L}^3}\sum_{n_\mu\neq 0}\frac{1}{2\left(i(\omega+\no)-{\Delta}\right)}\frac{(\nv)^2}{(\no)^2+(\nv)^2+m_{\pi}^2}\nn\\&\stackrel{\omega\rightarrow 0 }{\longrightarrow}&\frac{3\tilde{g}^2}{2f^2}{\mathbb A}(m_{\pi}, \Delta,L,\beta)=\frac{3\tilde{g}^2}{2f^2}{\mathbb A}(0, \Delta,L,\beta) +\mathcal{O}(\epsilon^5),\\  M^{(3)}(m_{K}, m_{\eta}, \Delta^s, L,\beta)&=&\frac{\tilde{g}^2}{f^2}{\mathbb A}(m_K, \Delta^s,L,\beta)+\frac{\tilde{g}^2}{6f^2}{\mathbb A}\left(m_{\eta}, \Delta^s,L,\beta\right) \end{eqnarray}
where $\tilde{g}=g+\frac{g_1}{m_{\text{Q}}}$. The first $\mathcal{O}(\epsilon^4)$ contribution comes from integrating out the zero-modes using Eq. (\ref{zero_modes}), depicted by Fig. (\ref{subleading}(a)),
\begin{eqnarray}
\label{delta2}
  M_a^{(4)}(m_{\pi}, L,\beta)&=& \frac{m_{\pi}^2}{2\Lambda_{\chi}}  (\sigma_1+2\sigma_2)\frac{X^\prime(s)}{X(s)}. 
\end{eqnarray}
The second graph, Fig. (\ref{subleading}(b)), corresponds to the four-point vertex contribution to the mass, 
\begin{eqnarray}
\label{delta3}
%  M_b^{(4)}(m_{\pi}, L,\beta)&=&frac{8\gamma_1+4\gamma_2}{2f^2\Lambda_{\chi}}{\mathbb B}(0,L,\beta) +\mathcal{O}(\epsilon^6),\nn\\
  M_b^{(4)}(m_{\pi}, L,\beta)&=&
-\frac{3\gamma_1+6\gamma_2}{2f^2\Lambda_{\chi}}{\mathbb B}(0,L,\beta) +\mathcal{O}(\epsilon^6),\nn\\
  M_b^{(4)}(m_{K},m_{\eta}, L,\beta)&=&
-\frac{4\gamma_1+\gamma_2}{f^2\Lambda_{\chi}}{\mathbb B}(m_K,L,\beta)
 -\frac{\gamma_3+4\gamma_4}{f^2\Lambda_{\chi}}\left({\mathbb B}(m_K,L,\beta) +{\mathbb C}(m_K,L,\beta)\right) \nn\\
&-&\frac{6\gamma_1+\gamma_2}{6f^2\Lambda_{\chi}}{\mathbb B}\left(m_{\eta},L,\beta\right)
-\frac{\gamma_3+6\gamma_4}{6f^2\Lambda_{\chi}}\left({\mathbb B}\left(m_{\eta},L,\beta\right)+{\mathbb C}\left(m_{\eta},L,\beta\right)\right).
%  M_b^{(4)}(m_{K},m_{\eta}, L,\beta)&=&\frac{8\gamma_1+2\gamma_2}{2f^2\Lambda_{\chi}}{\mathbb B}(m_K,L,\beta) +\frac{2\gamma_3}{2f^2\Lambda_{\chi}}\left({\mathbb B}(m_K,L,\beta) +{\mathbb C}(m_K,L,\beta)\right) \nn\\&+&\frac{6\gamma_1+2\gamma_2}{6f^2\Lambda_{\chi}}{\mathbb B}\left(m_{\eta},L,\beta\right)+\frac{2\gamma_3}{6f^2\Lambda_{\chi}}\left({\mathbb B}\left(m_{\eta},L,\beta\right)+{\mathbb C}\left(m_{\eta},L,\beta\right)\right) .
\end{eqnarray}
The third diagram, Fig. (\ref{subleading}(c)), comes from the three-point vertex corrections in the Lagrangian, and it results in the following contribution to the P meson mass:
\begin{eqnarray}
\label{delta4}
M_{c}^{(4)}(m_{\pi},\Delta, L,\beta)&=&-
6\frac{{g}g_2}{2f^2m_{\text{Q}}}{\mathbb A}(0, \Delta ,L,\beta)
+\mathcal{O}(\epsilon^5),\nn\\
M_{c}^{(4)}(m_{K}, m_{\eta}\Delta^s, L,\beta)&=&-
\frac{4{g}g_2}{2f^2m_{\text{Q}}}{\mathbb A}(m_K, \Delta^s,L,\beta)
-\frac{2{g}g_2}{6f^2m_{\text{Q}}}{\mathbb A}\left({\frac{2}{\sqrt{3}}}m_K, \Delta^s,L,\beta\right). 
\end{eqnarray}
In order to evaluate the temporal sum, the Abel-Plana formula will be used:
\begin{eqnarray}
  \label{finiteTtrick} \frac{1}{\beta}\sum_{n} f(\frac{2\pi n}{\beta}) =
\int_{-\infty}^\infty \frac{dz}{2\pi} f(z) - i \text{Res}
(\frac{f(z)}{e^{i\beta z}-1})|_{\rm lower plane} + i \text{Res}
(\frac{f(z)}{e^{-i\beta z}-1})|_{\rm upper plane}.
\label{abel}
\end{eqnarray}
The spatial sum can be performed using Poisson's 
Resummation formula,
\begin{eqnarray}
  \frac{1}{L^3}\sum_{\vec{n}} \frac{(\nv)^{2m}}{(\nv)^2+x^2} &=& \frac{1}{L^3}
\int d^3k \frac{k^{2m}}{k^2+x^2} \sum_{\vec{n}} \delta(\vec{k}-\frac{2\pi
  \vec{n}}{L}) = \int \frac{d^3k}{(2\pi)^3} \frac{k^{2m}}{k^2+x^2}
\underbrace{\sum_{\vec{n}} \delta(\frac{\vec{k}L}{2\pi}-\vec{n})
}_{\sum_{\vec{n}}e^{i L \vec{k}\cdot\vec{n}}}\nn\\
&=& \int \frac{d^3k}{(2\pi)^3} \frac{k^{2m}}{k^2+x^2} +
\frac{\left(-x^2\right)^m}{4\pi
  L}\sum_{\vec{n}\neq 0}\frac{e^{-nxL}}{n}.\label{eq:poisson}
\end{eqnarray}

 As carefully discussed in Ref. \cite{epsilon1}, the finite temperature contributions in the $\epsilon$-regime are usually heavily suppressed, and such is the case in all the integrals considered here. This allows one to safely neglect finite temperature terms. With this, one can extract the volume dependence of the sum in  Eq. (\ref{sum1}) as  
\begin{eqnarray}
\label{Aexact}
  \delta{\mathbb A} (m_{\text{l}},\Delta, L,\beta )&=& {\mathbb A} (m_{\text{l}},\Delta, L,\beta )-{\mathbb A} (m_{\text{l}},\Delta, L\rightarrow\infty,\beta\rightarrow\infty )\nn\\
&\stackrel{\omega\to 0}{\longrightarrow}&
    \Delta\int_{-\infty}^{\infty}\frac{dk_0}{2\pi}\frac{1}{k_0^2+\Delta^2} 
  \frac{k_0^2+m_{l}^2}{4\pi
  L}\sum_{\vec{n}\neq 0}\frac{e^{-n\sqrt{k_0^2+m_{l}^2}L}}{n} +\mathcal{O}(\epsilon^6).
% && +  \frac{1}{L^3}\sum_{\vec{n}}\frac{(\nv)^2}{e^{\beta\sqrt{(2\pi\vec{n}/L)^2+m_{l}^2}}-1 }\frac{\Delta}{(\nv)^2+m_{\pi}^2}  \frac{1}{\Delta^2+(\nv)^2+m_{l}^2}.
\end{eqnarray}

%%%%%%%%%%%%%%%%%%%%%%%%%%%%%%%
\subsection{1. $\epsilon$-Regime Integrals: $m_{\text{l}}=m_{\pi}$}
In the case that the sum arises from a pion loop, one can take the chiral limit and substitute $\sqrt{k_0^2+m_{l}^2}\rightarrow k_0$ in the above expression. Corrections to this approximation will result in $\mathcal{O}(\epsilon^6)$ contributions to the heavy meson masses.  All the integrals can be preformed using the following generating formula:
\begin{eqnarray}
\mathcal{I}(\alpha,\Delta)\equiv\int_0^{\infty}dk_0
\frac{1}{k_0^2+\Delta^2}e^{-k_0\alpha}
&=&\frac{\text{Ci}(\alpha\Delta)sin(\alpha\Delta)}{\Delta}
+\frac{\cos(\alpha\Delta)\left(\frac{\pi}{2}-\text{Si}(\alpha\Delta)\right)}{\Delta},
\label{generating1}\hspace{1cm}\\
\mathcal{I}''(\alpha,\Delta)\equiv\int_0^{\infty}dk_0
\frac{k_0^{2m}}{k_0^2+\Delta^2}e^{-k_0\alpha}
&=&\frac{\partial^{2m}}{\partial\alpha^{2m}}\mathcal{I}(\alpha,\Delta).
%\hspace{1cm}\int_0^{\infty}dk_0\frac{1}{\left(k_0^2+\Delta^2\right)^{m}}e^{-k_0\alpha}=\frac{(-1)^{m-1}}{(m-1)!}\frac{\partial^m}{\partial(\Delta^2)^m}\mathcal{I}(\alpha,\Delta).
\label{generating2}
\end{eqnarray}
%%%%%%%%%%%%%%%%%%%%%%%%%%%%%%%%%
%Since in all of these cases the zero modes, $n_{\mu}=(0, 0)$, do not contribute, one can expand in powers of $m_{\pi}L\sim\mathcal{O}(\epsilon)$, for example:\begin{eqnarray}\frac{1}{\beta \text{L}^3}\sum_{n_\mu\neq 0}  \frac{(\nv)^2}{(\no)^2+(\nv)^2+m_\pi^2}= \frac{1}{\beta \text{L}^3} \sum_{n_\mu\neq 0}  \frac{(\nv)^2}{(\no)^2+(\nv)^2}+\mathcal{O}(\epsilon^6),\end{eqnarray} This is by no means a proof but rather a justification for taking the chiral limit in the following calculations. It is convenient to take the chiral limit after performing the integrals, otherwise one would have to reintroduce it to regulate the IR divergences.

 Where $\text{Ci}(x)=\gamma+\log(x)+\int^x_0\frac{\cos(t)-1}{t}dt$ and $\text{Si}(x) =\int^x_0\frac{\sin(t)}{t}dt $ are the geometric integrals.  From Eq. (\ref{generating2}), it follows:
\begin{eqnarray}
\label{deltaA}
  \delta{\mathbb A} (0,\Delta,L,\beta )=&&
\frac{\Delta^2}{4\pi^2
  } \sum_{\vec{n}\neq 0}
  \frac{1}{nL\Delta}\left(\frac{1}{nL\Delta}
  - \text{Ci}(nL\Delta) \sin(nL\Delta)
+ \cos(nL\Delta)\left(\text{Si}(nL\Delta)-\frac{\pi}{2}\right)
\right).
\end{eqnarray}
Assuming $\Delta L\sim \epsilon$, it is possible to expand about $\Delta L= 0$. In this limit the sum may be approximated as an integral over the variable $z\equiv nL\Delta$, \cite{epsilon2},  
\begin{eqnarray}
L^2\Delta^2 \sum_{\vec{n}\neq 0}
  \frac{1}{nL\Delta}\left(\frac{1}{nL\Delta}
  - \text{Ci}(nL\Delta) \sin(nL\Delta)
+ \cos(nL\Delta)\left(\text{Si}(nL\Delta)-\frac{\pi}{2}\right)
\right)
\nn\\
\stackrel{L\Delta\to 0}{\longrightarrow}
4\pi\int_0^{\infty}\frac{z^2dz}{z}\left(\frac{1}{z}
  - \text{Ci}(z) \sin(z)
+ \cos(z)\left(\text{Si}(z)-\frac{\pi}{2}\right)
\right)=2\pi^2.
\end{eqnarray}
At leading order this matches to the approximation made in Ref. \cite{epsilon2} for the same integral. The next term in the expansion comes from taking a derivative with respect to $\alpha'\equiv L\Delta $
\begin{eqnarray}
\frac{\partial}{\partial\alpha'}\alpha'^2\sum_{\vec{n}\neq 0}
  \frac{1}{n\alpha'}\left.\left(\frac{1}{n\alpha'}
  - \text{Ci}(n\alpha') \sin(n\alpha')
+ \cos(n\alpha')\left(\text{Si}(n\alpha')-\frac{\pi}{2}\right)
\right)\right|_{\alpha'\rightarrow0}
=-\sum_{\vec{n}\neq 0}\frac{\pi}{2n}
\end{eqnarray}
Expanding about $L\Delta=0$ leads to a $\mathcal{O}(\epsilon^4)$ approximation of Eq. (\ref{deltaA})
\begin{eqnarray}
\label{Aepsilon}
 \delta {\mathbb A} (0,\Delta,L,\beta )=
\frac{1}{2L^3}
-\sum_{\vec{n}\neq 0}
{\frac{\Delta}{8n\pi L^2}}
+\mathcal{O}(\epsilon^5)=
\underbrace{\frac{1}{2L^3}}_{\mathcal{O}(\epsilon^3)}
-\underbrace{\frac{\Delta c_1}{8\pi L^2}}_{\mathcal{O}(\epsilon^4)}
+\mathcal{O}(\epsilon^5),
\end{eqnarray}
%%%%%%%%%%%%%%%%%%%%%%%%%%%%%%%%%
where Eq. (\ref{sums}) has been used. To $\mathcal{O}(\epsilon^4)$ the remaining integrals are 
\begin{eqnarray}
\label{Bepsilon}
 \delta{\mathbb B} (m_\pi, L,\beta )&=&
  \frac{1}{4\pi L}
  \sum_{\vec{n}\neq 0}\int_{-\infty}^{\infty} \frac{dk_0}{2\pi} k_0^2 \frac{e^{-nL\sqrt{k_0^2+m_{\pi}^2}}}{n}= \frac{1}{2\pi^2 \text{L}^4}
  \sum_{\vec{n}\neq 0}  \frac{1}{n^4}+\mathcal{O}(\epsilon^6)= 
  \frac{c_4}{2\pi^2 \text{L}^4} +\mathcal{O}(\epsilon^6),\\
  \label{Cepsilon}
 \delta{\mathbb C} (m_\pi, L,\beta )&=&
-\frac{1}{2\pi^2
  \text{L}^4} \sum_{\vec{n}\neq 0}
  \frac{1}{n^4}+\mathcal{O}(\epsilon^6)= 
 - \frac{c_4}{2\pi^2 \text{L}^4} +\mathcal{O}(\epsilon^6).
   \end{eqnarray}
In writing out the full expression of the masses, it is important to note that $\delta{\mathbb B} (0, L,\beta )+\delta{\mathbb C} (0, L,\beta )=0+\mathcal{O}(\epsilon^6)$. Two previously calculated sums have been used \cite{sum1, lusche2, sum3, sum4}:
\begin{eqnarray}
\label{sums}
 c_1=\sum_{\vec{n}\neq 0}
  \frac{1}{|n|}= -2.8372974\hspace{1cm}
c_4=\sum_{\vec{n}\neq 0}
  \frac{1}{|n|^4}=16.532315.
 \end{eqnarray}
 \vspace{-1cm}
%%%%%%%%%%%%%%%%%%%%%%%%%%%%%%%%%%%%
\subsection{2. p-Regime Integrals: $m_{\text{l}}=\{m_{K},m_{\eta}\}$}
In the p-regime, the light meson mass is comparable to the lowest non-zero momentum $m_{\text{l}}/\Lambda_{\chi}\sim2\pi/L\Lambda_{\chi}\sim\epsilon$. In this regime the small mass approximations used in the previous sections are no longer valid. One must perform the integral in Eq. (\ref{Aexact}) without taking the chiral limit. Although this integral cannot be evaluated exactly, in the $\Delta\rightarrow 0$ limit the integral is dominated by small values of $k_0$. In this case, the argument in the exponential can be approximated as $\sqrt{k_0^2+m_{l}^2}=m_{l}+\frac{k_0^2}{2m_{\text{l}}}-\frac{k_0^4}{8m_{\text{l}}^3}+\cdots$,
\begin{eqnarray}
\Rightarrow \delta{\mathbb A} (m_{\text{l}},\Delta, L,\beta )&=&\Delta\sum_{\vec{n}\neq 0}\int_{-\infty}^{\infty}\frac{dk_0}{2\pi}\frac{1}{k_0^2+\Delta^2} 
  \frac{k_0^2+m_{l}^2}{4\pi
  L}\frac{e^{-nm_{l}L}e^{-\frac{nLk_0^2}{2m_{l}}}}{n} \left(1+\frac{nLk_0^4}{8m_{\text{l}}^3}\right)+\cdots \nn\\
  &=&\frac{m_{\text{l}}^2}{8\pi L}\sum_{\vec{n}\neq 0}\frac{e^{-nm_{l}L}}{n}+\sum_{\vec{n}\neq 0}\left(\frac{3+9m_{\text{l}}nL-(m_{\text{l}}nL)^2}{64\pi^2m_{\text{l}}^{1/2}L^{5/2}n^{5/2}}\right)\sqrt{2\pi}\Delta{e^{-nm_{l}L}} + \mathcal{O}(\epsilon^5)\nn\\
 & \equiv&\frac{m_{\text{l}}^2}{8\pi L}\mathcal{N}_1(m_{\text{l}}, L)+
\frac{\Delta}{m_{\text{l}}^{1/2}L^{5/2}} \mathcal{N}_2(m_{\text{l}}, L)+ \mathcal{O}(\epsilon^5),
\label{Ap}
  \end{eqnarray}
  where the definition in Eq. (\ref{nfnctions}) have been used. The remaining integrals can be performed exactly,   
\begin{eqnarray}
 \delta{\mathbb B} (m_{\text{l}}, L,\beta )&=&
  \frac{1}{4\pi L}
  \sum_{\vec{n}\neq 0}\int_{-\infty}^{\infty} \frac{dk_0}{2\pi} k_0^2 \frac{e^{-nL\sqrt{k_0^2+m_{l}^2}}}{n}=
   \frac{m_{\text{l}}^3}{16\pi^2 L}
  \sum_{\vec{n}\neq 0} \left(K_3(m_{\text{l}}nL)-K_1(m_{\text{l}}nL)\right)+\mathcal{O}(\epsilon^5),\nn\\
  \label{Bp}
  &\equiv&  \frac{m_{\text{l}}^3}{16\pi^2 L}\mathcal{K}_1(m_{\text{l}}L)+\mathcal{O}(\epsilon^5)\label{Bp}\\
\delta{\mathbb C} (m_{\text{l}}, L,\beta )&=&
  \label{Cp}
- \frac{m_{\text{l}}^3}{16\pi^2 L}
  \sum_{\vec{n}\neq 0} \left(K_3(m_{\text{l}}nL)+3K_1(m_{\text{l}}nL)\right)+\mathcal{O}(\epsilon^5)\equiv - \frac{m_{\text{l}}^3}{16\pi^2 L}\left(\mathcal{K}_1(m_{\text{l}}L)-\mathcal{K}_2(m_{\text{l}}L)\right) \label{Cp}.
   \end{eqnarray}
In writing  Eqs. (\ref{Ap}-\ref{Cp}) the following dimensionless functions were used,  
 \begin{eqnarray}
\label{nfnctions}
  \mathcal{N}_1(m_{\text{l}}, L)=\sum_{\vec{n}\neq 0}\frac{e^{-nm_{l}L}}{n}&\hspace{1cm}&
  \mathcal{N}_2(m_{\text{l}}, L)=\sum_{\vec{n}\neq 0}\left(\frac{3+9m_{\text{l}}nL-(m_{\text{l}}nL)^2}{64\pi^2n^{5/2}}\right)\sqrt{2\pi}{e^{-nm_{l}L}} \\
\label{kfnctions}  \mathcal{K}_1(m_{\text{l}}L)= \sum_{\vec{n}\neq 0} \left(K_3(m_{\text{l}}nL)-K_1(m_{\text{l}}nL)\right)&\hspace{1cm}&
  \mathcal{K}_2(m_{\text{l}}L)= -4 \sum_{\vec{n}\neq 0} K_1(m_{\text{l}}nL),
  \end{eqnarray}
where $K_{\alpha}$ are the modified  Bessel functions of the second kind. Finally, by adding the contributions from Eqs. (\ref{delta1}-\ref{delta4}) and substituting the expressions of the respective sums, one arrives at Eq. (\ref{results1}).
   
\bibliography{bibi}
\end{document}